\documentclass[conference]{IEEEtran}

\IEEEoverridecommandlockouts

\usepackage[utf8]{inputenc}
\usepackage[T1]{fontenc}
\usepackage{graphicx}
\usepackage{xcolor}
\usepackage{amsmath,amssymb}
\usepackage{braket}
\usepackage{bm}
\usepackage{booktabs}
\usepackage{hyperref}
\usepackage{tabularx}
\usepackage{tikz}
\usetikzlibrary{patterns,positioning,arrows.meta,shapes.misc,fit,calc}
\usepackage[dvipsnames,svgnames,x11names]{xcolor}

\usepackage{afterpage}

\hypersetup{
    colorlinks=true,
    linkcolor=blue,
    citecolor=blue,
    urlcolor=blue
}

\newcommand{\Graph}{{\mathcal{G}}}
\newcommand{\edges}{{\mathcal{E}}}
\newcommand{\vertices}{{\mathcal{V}}}
\newcommand{\weights}{{\mathcal{W}}}

\title{\LARGE \bf A Scalable Heuristic for Molecular Docking on Neutral-Atom Quantum Processors}

\author{Mathieu Garrigues$^\dagger$, Victor Onofre$^\dagger$, Wesley Coelho, and S. Acheche%
\thanks{$^\dagger$ These authors contributed equally to this work.}
\thanks{All authors are with Pasqal, 24 rue Emile Baudot, 91120 Palaiseau, France.}%
}

\begin{document}

\maketitle
\thispagestyle{plain}
\pagestyle{plain}


\begin{abstract}
Molecular docking is a critical computational method in drug discovery used to predict the binding conformation and orientation of a ligand within a protein's binding site. Mapping this challenge onto a graph-based problem, specifically the Maximum Weighted Independent Set (MWIS) problem, allows it to be addressed by specialized hardware such as neutral-atom quantum processors. However, a significant bottleneck has been the size mismatch between biologically relevant molecular systems and the limited capacity of near-term quantum devices.
In this work, we overcome this scaling limitation by the use of a divide-and-conquer heuristic introduced in Cazals \textit{et al}~\cite{cazals2025gadgets}. This algorithm decomposes a single, intractable graph instance into smaller sub-problems that can be solved sequentially on a neutral-atom quantum emulator, incurring only a linear computational overhead. We benchmark this approach on 10 real-world protein-ligand complexes, including 9 from the Astex Diverse Set, with graphs ranging from 225 to 585 vertices. The quantum heuristic consistently outperforms a greedy baseline and achieves the provably optimal solution on a 540-node instance (TACE-AS). We further assess the biological relevance of the reconstructed poses via the fraction of native contacts, and benchmark the full workflow on a standard dataset of diverse protein-ligand complexes.
Our work establishes a scalable blueprint for applying quantum optimization to molecular docking, while identifying concrete directions for improving both the algorithmic strategy and the underlying graph model.

\end{abstract}

\section{INTRODUCTION}
\afterpage{
\begin{figure*}[t]
\includegraphics[width=\textwidth, trim=.74cm 1.32cm .74cm .74cm, clip]{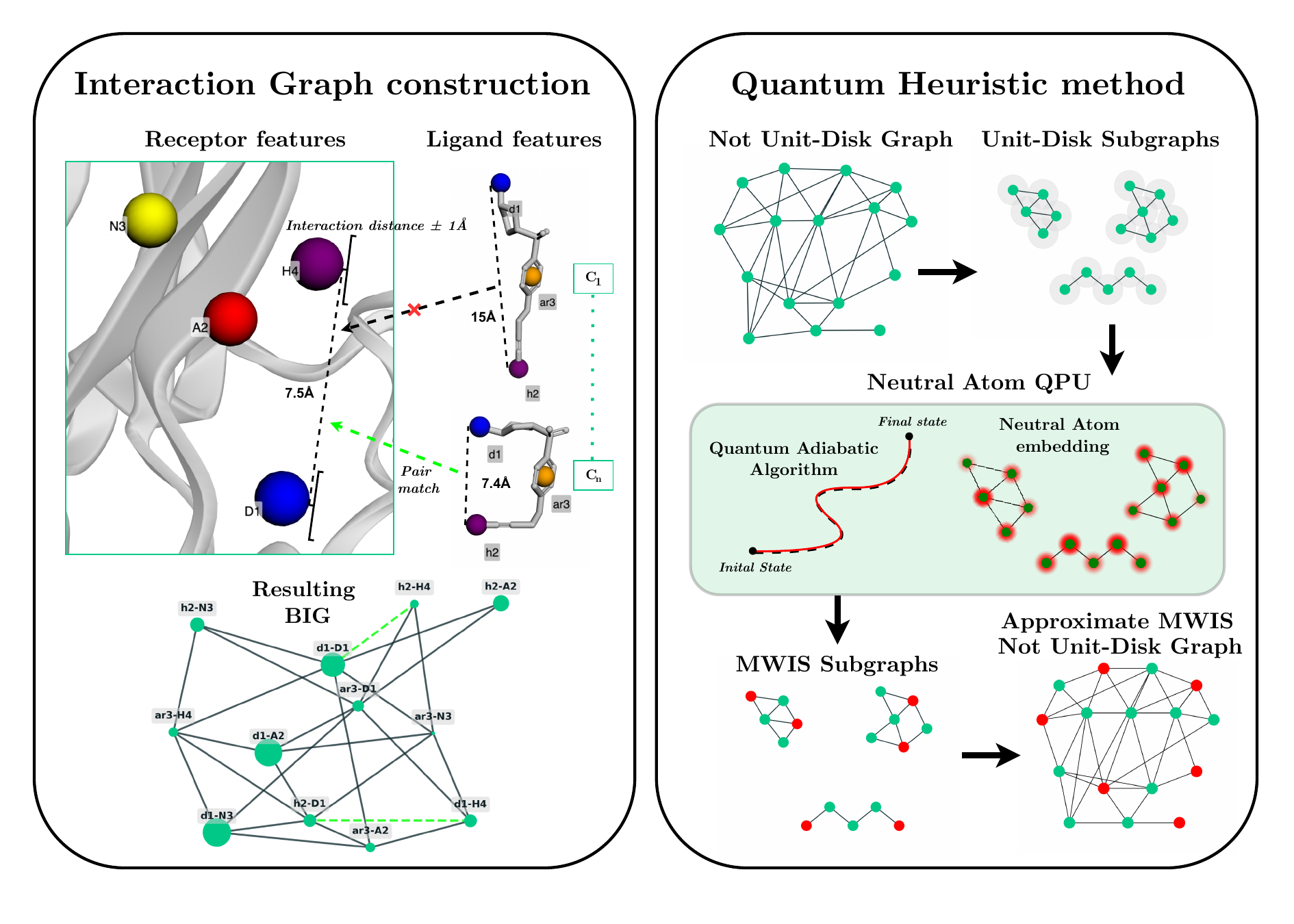}
\caption{
\textbf{(left)} Pharmacophore features are extracted from receptor
and ligand (\ref{sec:baseline_method}), displayed as colored spheres.
Ligand conformers $\{C_1,\ldots,C_n\}$ with varying geometries are
generated (\ref{sec:imp_flexibility}). Each BIG node represents a
possible receptor–ligand pharmacophore contact pair. Two nodes are
connected if geometric compatibility is satisfied in at least one
conformer (eq.~\ref{eq:flex_edge}, green dashed arrow). The Maximum
Weighted Clique (MWC) of the resulting non-Unit-Disk (UD) graph
yields the strongest interaction ensemble.
\textbf{(right)} Via graph complement duality, the MWC maps to a
Maximum Weighted Independent Set (MWIS). The Quantum Heuristic
decomposes the non-UD BIG into UD subgraphs embeddable on the QPU,
where local detuning encodes vertex weights. A Quantum Adiabatic
protocol solves the MWIS on each
subgraph~\cite{cazals2025gadgets}; partial solutions are
combined to approximate the full-graph MWIS. The resulting interaction
set specifies receptor points $p_i$ and ligand counterparts $l_i$
used to reconstruct the docking pose.}
\label{fig:overview}
\end{figure*}
}
Molecular docking is a fundamental tool in computational drug discovery, aiming to predict the binding conformation of a ligand within a protein's binding site~\cite{Guedes2021}. The accuracy and efficiency of this process are critical for the high-throughput screening of vast chemical libraries. The field is dominated by established computational workhorses, primarily force-field-based methods, which perform a stochastic search of the conformational space~\cite{autodock, glide}. Alongside these, emerging paradigms such as deep learning are showing promise~\cite{Shen2023}. The immense computational cost of these simulations remains a significant bottleneck, motivating the exploration of entirely new formalisms and computing platforms.

An alternative formalism is to translate the continuous and complex problem of molecular interactions into a discrete combinatorial optimization problem. This is achieved by abstracting the system into a graph. In this approach, potential favorable interactions between pharmacophore points on the ligand and the receptor (protein) are represented as vertices. The core challenge then becomes finding the best possible subset of these interactions that are all mutually compatible. This can be formally mapped to the Maximum Weighted Independent Set (MWIS) problem. The goal of the MWIS problem is to find a set of vertices in a graph that have no edges between them, such that the sum of their weights is maximized. This elegant mapping provides a structured, discrete representation of the docking challenge.

While conceptually powerful, solving the MWIS problem is NP-Hard, making it intractable (on the worst-case scenario) for classical computers on large, densely connected graphs~\cite{cazals2025identifyinghardnativeinstances}. Remarkably, however, this problem has a natural physical analogue in a neutral-atom Quantum Processing Unit (QPU). On these devices, the MWIS problem can be directly embedded by arranging atoms according to the graph's structure, where the Rydberg blockade mechanism, a phenomenon preventing the simultaneous excitation of nearby atoms, provides a physical manifestation of the independent set constraint~\cite{pichler2018quantum}. However, a major obstacle in applying this graph-based quantum approach has been the significant size mismatch between biologically relevant problems and the limited scale of near-term (NISQ) quantum hardware. Past approaches~\cite{molecular_yu_shang, banchi2020molecular, lancellotti2024experimental, ding2024molecular}, including our own~\cite{garrigues2024moleculardockingneutralatoms}, were therefore constrained to using highly simplified graph models. These models were forced to drastically reduce the number of considered interactions to ensure the resulting graph could be embedded on a QPU, which fundamentally limited their predictive power and physical realism.

In this paper, we introduce a complete, end-to-end workflow to solve the molecular docking problem on a neutral-atom QPU, designed from the ground up to overcome the aforementioned scaling limitations. Its crucial parts are presented in Fig.~\ref{fig:overview}. The core of our contribution is a methodology that synergistically combines a physically-aware graph construction protocol with a scalable solver. Our main objective is to establish this workflow as a viable proof of concept for tackling the abstract graph-based formulation of docking at a biologically relevant scale, a critical first step in evaluating the potential of this computational paradigm. By implementing a divide-and-conquer heuristic introduced in Cazals \textit{et al}~\cite{cazals2025gadgets}, the algorithm can decompose a single, large graph instance into a series of smaller, tractable sub-problems that are solved sequentially on a quantum emulator. This removes the critical size barrier and enables the use of a graph model that is sufficiently complex to be more physically meaningful.

Our integrated workflow consists of three main contributions: (1) A physically-aware graph construction protocol that incorporates an expanded interaction scope, realistic ligand flexibility via conformational ensembles, and solvent-accessibility filtering. (2) The implementation of a decomposition-based heuristic that enables solving graphs of a size previously inaccessible to quantum approaches. (3) A comprehensive demonstration of this entire pipeline on 10 real world biochemical systems, with obtaining of an optimal solution for the TACE-AS complex, formulated as a 540-node MWIS problem.
The code, graph instances and results for this work are publicly available~\cite{garrigues2025moleculardocking}.

\section{MOLECULAR DOCKING}
\label{sec:reduction_drawbacks}

\subsection{Overview of the Baseline Binding Interaction Graph Docking Method}
\label{sec:baseline_method}

Our work builds upon and extends the graph-based molecular docking approach introduced by Bianchi \textit{et al}.~\cite{banchi2020molecular}. Their method transforms the geometric challenge of docking into a well-known combinatorial optimization task: the Maximum Weighted Clique (MWC). This problem is the task of finding a clique, a subset of vertices where every vertex is connected to every other, such that the sum of the weights of the vertices in the clique is maximized.

The core principle is the construction of a single Binding Interaction Graph (BIG), denoted as $\Graph = (\vertices, \edges, \weights)$, where vertices $\vertices$ and edges $\edges$ encode interactions between a ligand and a protein's binding pocket. Vertices weights $\weights = \{w_1,\ldots, w_{|\vertices|}\}$ represent the strength of the interaction. The construction of this graph follows several key steps:

\begin{enumerate}
\item \textbf{Pharmacophore Representation:} The ligand and protein
are first simplified into a set of key chemical interest points,
or pharmacophores. We leverage the \texttt{RDKit}
library~\cite{rdkit_zenodo}, which provides a robust framework for
feature detection based on SMARTS (SMiles ARbitrary Target
Specification) patterns. The methodology identifies six
primary feature families essential for describing molecular
recognition events: Negative Ionizable (NI), Positive
Ionizable (PI), Hydrogen Bond Donor (D), Acceptor (A),
Hydrophobic (H), and Aromatic (AR). Each detected feature
may involve one or more atoms and is assigned to one of
these families based on its chemical role.

\item \textbf{Vertex Definition:} Each vertex $v \in \vertices$ in the graph $\Graph$ does not represent a single pharmacophore but rather an interaction pair. Specifically, a vertex $v$ corresponds to the pairing of a ligand pharmacophore $l$ with a protein pharmacophore $p$.
The interaction has a certain strength, depending on the families of the two considered pharmacophores. Crucially, their mutual distance is not taken into account, and the interaction is considered as either happening or not. It is represented by the corresponding vertex's weight $w(v)$ given by:
\begin{equation}
w(v) = \mathcal{P}(\text{Family}(l), \text{Family}(p))
\label{eq:vertex_weight}
\end{equation}

where:
\begin{itemize}
\item $v = (l, p)$ is the vertex representing the interaction pair.
\item $\text{Family}(x)$ is a function that returns the pharmacophore family (e.g., NI, PI, D, A, H, AR) of a point $x \in l \uplus p$.
\item $\mathcal{P}(f_1, f_2)$ is the knowledge-based potential function that returns the interaction score between family $f_1$ and family $f_2$. Those potentials are available in the source code for this work \cite{garrigues2025moleculardocking}.
\end{itemize}

\item \textbf{Edge Definition:} An edge is established between two vertices, $v_a = (l, p)$ and $v_b = (l', p')$, if and only if the two interactions they represent are geometrically compatible. This compatibility is ensured if the simultaneous activation of both interactions preserves the ligand's rigid internal structure. This condition is met if the distance between the two protein pharmacophores, $d(p, p')$, is similar to the euclidean distance between the two corresponding ligand pharmacophores, $d(l, l')$, within a given tolerance.
\end{enumerate}

Once this BIG is constructed, the docking problem is redefined as finding the MWC of this graph. The MWC therefore corresponds to the largest set of mutually compatible ligand-protein interactions, which in turn defines the final docking pose. For a comprehensive description of the method, including the initial discretization of ligand poses, we refer the reader to the original publication~\cite{banchi2020molecular}.

While this approach is effective in transforming a continuous problem into a discrete structure, its binary nature (an interaction is either compatible or not) and purely geometric foundation simplify the complexity of molecular interactions. Mainly, actual ligand flexibility is not taken into account, and points which are physically inaccessible to the ligand can be in the final solution. In the following sections, we describe three significant improvements to this graph model aimed at incorporating a more faithful physical and chemical representation.

\subsection{Improvements to the Interaction Graph}
\label{sec:improvements}

To enhance the physical and chemical realism of the baseline model, we introduce three key modifications to the graph construction process. These improvements are designed to create a more accurate and descriptive problem representation.

\subsubsection{Enriching the Interaction Set}
\label{sec:imp_interactions}

Our approach modifies the graph construction process in two main aspects to provide a more exhaustive model of the binding site.

First, the spatial scope for selecting receptor pharmacophores is expanded by including all receptor pharmacophore points located within a 6.5~\AA~ (against 4~\AA~ in previous works) radius of any atom of the ligand in its initial pose. In this experiment, the initial pose is the ligand's known crystallographic pose in the complex. The objective of this larger cutoff is to ensure that the entire binding site and its immediate chemical context are considered in the model.

Second, we alter the method for vertex generation. Instead of creating vertices based on a pre-filtered set of interaction pairs, our method adopts an all-pairs approach. Let $L_{pharma}$ be the set of pharmacophores on the ligand and $P_{pharma}$ be the set of receptor pharmacophores identified within the expanded spatial scope. The vertex set $\vertices$ is then constructed from the cartesian product of these two sets, such that a vertex is created for every possible pairing. The total number of vertices is therefore $|\vertices| = |L_{pharma}| \times |P_{pharma}|$. Each vertex is subsequently assigned a weight $w(v)$ based on the interaction potential between the corresponding pharmacophore families, as defined in Equation~\ref{eq:vertex_weight}.

These modifications result in a larger and more densely connected interaction graph. This approach increases the size of the optimization problem, as the task of identifying the most relevant interactions is shifted from the graph construction heuristics to the solver itself. The resulting model provides a more complete inventory of the possible interactions within the binding site.

\subsubsection{Filtering by Solvent-Accessible Surface Area (SASA)}
\label{sec:imp_sasa}

The initial identification of pharmacophores on the receptor generates a comprehensive but noisy set of potential interaction points. A significant number of these points are located on atoms buried deep within the protein's core, making them sterically inaccessible to a binding ligand. Including these physically unreachable points in the graph construction has two major detrimental effects: it unnecessarily increases the size of the graph and the complexity of the optimization problem, and it introduces a significant amount of computational noise that can mislead the solver towards physically impossible solutions.

To address this, we introduce a critical pre-processing step to filter out these inaccessible points using the Solvent-Accessible Surface Area (SASA) as a metric. SASA quantifies the surface area of an atom that is accessible to a solvent molecule. It is typically calculated using a `rolling ball' algorithm like the one described by Shrake \& Rupley \cite{Shrake1973}, where a probe sphere of a given radius (approximating a water molecule) is rolled over the protein's van der Waals surface. Atoms with a high SASA value are exposed on the protein's surface, while atoms with a low or zero SASA value are considered buried.

For each protein structure, we compute the SASA using the Shrake-Rupley algorithm as implemented in \texttt{Biopython} \cite{biopython}. A pharmacophore point is retained for the graph construction only if the SASA of its parent atom exceeds a defined threshold, $\tau$. We set $\tau = 1.0$~\AA$^2$, a value slightly above zero to robustly exclude atoms that are almost completely buried but might have negligible surface exposure due to minor crevices.

\subsubsection{Incorporating Ligand Flexibility via Conformational Ensembles}
\label{sec:imp_flexibility}

To account for ligand flexibility, Bianchi \textit{et al}'s model applies a uniform, fixed tolerance to all intramolecular distances within the ligand. While straightforward to implement, this approach does not realistically reflect the ligand's structure, as the actual flexibility between any two points is highly dependent on the underlying bond connectivity and stereochemistry.

Our method employs a more physically grounded approach. We explicitly model flexibility by generating a representative ensemble of low-energy ligand conformers using \texttt{RDKit}. This ensemble implicitly defines a set of structurally plausible intramolecular distances, replacing the uniform flexibility assumption with a data-driven model based on the ligand's specific conformational landscape.

The core modification resides in the definition of an edge within the compatibility graph, which is now determined by the entire conformational ensemble. An edge is created if the geometric compatibility constraint is satisfied in \textit{at least one} of the generated conformers. Formally, an edge $(v_a, v_b) \in \edges$ exists if the following condition holds:
\begin{equation}
\exists C_n \quad \text{s.t.} \quad |d_{C_n}(l, l') - d(p, p')| \le 2\epsilon
\label{eq:flex_edge}
\end{equation}

where:
\begin{itemize}
\item $v_a = (l, p)$ and $v_b = (l', p')$ are two vertices in the graph.
\item $C_n$ is a specific conformer from the ligand's conformational ensemble.
\item $d_{C_n}(l, l')$ is the Euclidean distance between ligand pharmacophores $l$ and $l'$ in conformer $C_n$.
\item $d(p, p')$ is the distance between the corresponding protein pharmacophores.
\item $\epsilon$ is a fixed parameter representing a characteristic interaction distance.
\end{itemize}

The parameter $\epsilon$ is set to 1.0~\AA~ in our model. This value serves as a proxy for the effective radius of a pharmacophore interaction. The resulting tolerance of $2\epsilon$ on the distance difference accommodates the combined positional variance of the two pharmacophores involved in the interaction pair. This ensemble-based method effectively creates a physically-aware graph that embeds the relevant conformational space of the ligand, allowing the solver to identify poses that may rely on non-ground-state ligand geometries. A key limitation of this ensemble-based approach is its potential to generate `chimeric' cliques. These combine interactions sourced from distinct conformers, meaning no single conformer can simultaneously satisfy all the interactions within the resulting set. Post-processing steps may be necessary in order to choose the most fitting conformer in that case.

\subsection{BIG to a MWIS Problem}
\label{sec:big_to_mwis}

The graph construction process described previously results in a weighted graph, $\Graph = (\vertices, \edges, \mathcal{W})$, where the optimal docking pose corresponds to the solution of the MWC problem.

While MWC is a direct formulation, we transform it into the equivalent Maximum Weighted Independent Set (MWIS) problem. As a reminder, it is the problem of finding a set of vertices in a graph, no two of which are adjacent, that maximizes the total sum of their assigned weights.

This choice is motivated by the target quantum computing architecture. The MWIS problem is particularly well-suited for implementation on a neutral-atom QPU~\cite{ebadi2022quantum, pichler2018quantum, byun2022quantum}. On these devices, the problem can be embedded by arranging atoms in a configuration defined by the graph's vertices. The Rydberg blockade mechanism then provides a physical constraint that is a direct analogue to the mathematical constraint of an independent set: two atoms within a certain blockade radius cannot be simultaneously excited to the Rydberg state, which mirrors the rule that two connected vertices cannot both belong to an independent set~\cite{pichler2018quantum}.

The formal transformation is achieved by constructing the complement graph, denoted $\Graph_c = (\vertices, \edges_c, \weights)$. This graph shares the same vertex set $\vertices$ and vertex weights as the original BIG. The edge set, however, is inverted: an edge $(v_a, v_b)$ exists in $\Graph_c$ if and only if it does \textit{not} exist in $\Graph$.

By this construction, a clique in the original graph $\Graph$ is, by definition, an independent set in the complement graph $\Graph_c$. The problem of finding the MWC in $\Graph$ is therefore formally equivalent to finding the MWIS in $\Graph_c$.

All subsequent sections will address the problem of solving this MWIS formulation of the docking challenge.

\section{QUANTUM APPROACH}

\subsection{MWIS Solving with Neutral Atoms}
\label{sub:quantum_annealing}

Given a graph $\Graph = (\vertices, \edges, \weights)$, the MWIS problem is to find a subset $S \subseteq \vertices$ such that no two vertices in $S$ are adjacent, and the total weight $\sum_{i \in S} w_i$ is maximized. The problem can be formulated as the minimization of cost function,
\begin{equation}
\label{cost_function_mwis}
\min_{x \in \{0,1\}^n} \left( - \sum_{i=1}^n w_ix_i + \alpha \sum_{(i,j)\in \edges} x_i x_j \right),
\end{equation}

where $x_i$ is a binary variable, with 1 if vertex $i$ is included in the set and 0 if vertex $i$ is not included. The value of $\alpha$ must be large enough to ensure that violating a constraint is always worse than any potential gain in weight from making that violation.

In a neutral atom QPU of single $^{87}$Rb atoms trapped in arrays of optical tweezers~\cite{browaeys2020many,henriet2020quantum,Morgado2021}, qubits are encoded in the atomic ground state $\ket{g}$ and the Rydberg state $\ket{r}$. The dynamics of $N$ qubits are governed by the following Hamiltonian:
\begin{equation}
\label{eq:hamiltonian}
\begin{split}
\frac{H(t)}{\hbar} = & \sum_{i=1}^N \Big(\frac{\Omega(t)}{2} \big( |g\rangle_i\langle r|_i  +  |r\rangle_i\langle g|_i \big)\\
& - \delta_i(t) \hat{n}_i +  \sum_{j<i}\frac{C_6}{\hbar R_{ij}^6} \hat{n}_i \hat{n}_j \Big)
\end{split}
\end{equation}
where it describes the effect of a pulse on two energy levels of an individual atom, $|r\rangle$ and $|g\rangle$. A pulse is determined by its duration $\Delta t$, its Rabi frequency $\Omega(t)$, its detuning $\delta(t)$, between $0$ and $t$, where we call local detuning to the individual $\delta_{i}(t)$ associated to each qubit. With $\hat n_i=\ket{r}_i\bra{r}_i$ as the number operator, the distance between the atoms $R_{ij}$ and $C_6$, the Ising interaction coefficient depending on the Rydberg state considered~\cite{Beguin_2013}.

Given the neutral atoms properties, specifically the Rydberg blockade mechanism~\cite{bermot2025rydberg}, the MWIS cost function can be encoded on the Rydberg Hamiltonian as:
\begin{equation}\label{eq:operator_cost_function}
C_\Graph(\hat n)=\hat C_\Graph=-\sum_{i\in\vertices}w_i\hat n_i+\alpha \sum_{i<j}\hat n_i \hat n_j.
\end{equation}
This formulation is equivalent to the classical MWIS problem from eq.~\ref{cost_function_mwis}, where the binary variable $x_i$ is replaced by the number operator $\hat n_i$. Nodes are weighted by continuous values, $w_i\in(0,1)$, represented by taking a local detuning $\delta_i \propto  w_i$. The graph is embedded in a 2D lattice such that the nearest-neighbors interaction $C_{6}/{R_{ij}^6}$ encodes the uniform edge weight $\alpha$.

The native graphs that can be embedded in this neutral-atom machine are the Unit-Disk (UD) graphs. Such a graph can be embedded such that two vertices are connected by an edge if and only if they are separated by a distance smaller than a unit radius.

To find the MWIS in a given $\Graph$, the quantum state is initialized as $\ket{\psi}=\ket{0}^{\otimes N}$ and the QPU Hamiltonian (eq.~\ref{eq:hamiltonian}) is slowly annealed towards the Hamiltonian (eq.~\ref{eq:operator_cost_function}), driving the initial state towards the final ground state of the latter~\cite{farhi2014quantum}. A ground state of the cost function Hamiltonian $\hat C_\Graph$ is an optimal solution for eq.~\ref{cost_function_mwis}.
The time-dependence of the control fields ($\Omega(t)$ and $\delta(t)$) is known as the annealing schedule. In this work, we adopt the annealing schedule used in~\cite{cazals2025gadgets} and~\cite{cazals2025identifyinghardnativeinstances}.

The initial and final values of $\Omega$ are set such that the initial Hamiltonian ground state corresponds to the prepared initial state, and the final Hamiltonian encodes the optimization problem. We represent vertex weights using local detuning~\cite{goswami2024solving}, as shown in eq.~\ref{eq:hamiltonian}, which can be implemented in the QPU with an additional light-shifting potential from a Spatial Light Modulator (SLM). This allows programmable relative weighting~\cite{local_detuning}.

By arranging the qubits and executing the sequence, the quantum system is prepared in a state that encodes the solution to the MWIS problem. Sampling this quantum superposition $n_{\rm shots}$ times allows for measuring all degenerate MWIS solutions of $\Graph$, returning a distribution of bitstrings that represent either perfect (MWIS) or approximate (where the solution contains $v$ fewer vertices than the optimal one). Non-independent-set solutions are filtered out in post-processing if necessary.
However, since our BIG is not a UD graph, a new embedding method is necessary.

\subsection{Quantum Heuristic Method}
\label{sec:heuristics}

We address the MWIS problem of non-UD graphs using a heuristic approach developed in Cazals \textit{et al}~\cite{cazals2025gadgets}. The core of this method is a recursive algorithm that computes a solution for a general graph $\Graph$ by merging solutions from embeddable UD subgraphs, denoted as $\Graph'_i$.

To create valid subgraphs from a general graph $\Graph$ on a neutral atom pre-calibrated lattice $L$, we use a Greedy Lattice Subgraph (GLS) mapping~\cite{cazals2025gadgets}. It begins by selecting a random trap of $L$ and assigning it to an initial node in the input graph $\Graph$. The algorithm expands the mapped region by exploring the unmapped neighbors of the nodes that have already been placed, attempting to assign them to nearby, unoccupied lattice sites. Any neighbors that cannot be placed in the current iteration are discarded, as they will not find valid positions later in the process.

To solve the MWIS problem on a graph $\Graph$, we follow an iterative process. At each iteration $i$, we extract a list of embeddable subgraphs $\Graph'_{i}$, using the GLS mapping, choosing the one with the most vertices and solve the MWIS problem on it, saving the resulting independent set. To ensure the independence of the final resulting set, we then remove the selected vertices and their neighborhoods from the graph $\Graph_{i}$. This process is repeated until the remaining graph is empty, at which point we store the final approximate solution. The independent set generated through this method is always maximal.

In practice, at each iteration of the statevector-emulator-based simulation, we sample $k$ embeddable subgraphs from the current graph and solve the MWIS problem on each independently. From every subgraph solution, we retain the $s$ highest-weight independent sets. For the quantum solver, these correspond to the $s$ most probable bitstrings in the output distribution. Ideally, the highest-probability bitstring coincides with the MWIS, but the adiabatic evolution may also prepare sub-optimal solutions in practice, partially because the same pulse schedule is shared across all graphs. In a `breadth-first search' manner~\cite{Moore}, we repeat this process, decomposing each surviving graph further and solving the MWIS problem. To avoid the exponential growth of the process we trim back to the best $\ell$ branches at each level. Merging the largest independent sets found across all branches as the final MWIS. For a more detailed account of the quantum heuristic method we refer the reader to the original publication in~\cite{cazals2025gadgets}.

\addtolength{\textheight}{-3cm}


\section{RESULTS}

To evaluate the scalability and generality of our approach, we benchmark on 10 protein-ligand complexes: 9 drawn from the Astex Diverse Set~\cite{hartshorn2007}, a widely adopted benchmark for docking validation comprising 85 high-quality, drug-like complexes across diverse therapeutic target families, and the TACE–aryl sulfonamide complex (PDB: 2oi0) used in prior works.
The numerical simulations were performed using the Quantum Heuristic method described in Section~\ref{sec:heuristics}. We compared the performance of our method against three classical baselines: the state-of-the-art exact solver \texttt{CPLEX}~\cite{cplex2024}, Simulated Annealing~\cite{Garzillo_pasqal_qubo-solver_2025}, and a standard greedy heuristic for the MWIS problem~\cite{sakai2003greedy}. This algorithm constructs a maximal independent set by iteratively selecting the vertex that maximizes the weight-to-degree ratio $w_v/d(v)$ with $d(v)$ the degree of the vertex, adding it to the solution and removing it along with its neighborhood from the graph. This procedure guarantees a solution weight of at least $\sum_{v \in \mathcal{V}} \frac{w_v}{d(v) + 1}$.

\subsection{Optimization Performance}

Following the methodology detailed in Sections~\ref{sec:baseline_method} and~\ref{sec:improvements}, we constructed BIGs for all 10 complexes. The resulting complement graphs range from 225 to 585 vertices and from 17,806 to 114,429 edges, scales that were previously considered intractable for direct embedding on quantum hardware. Each subgraph instance was solved with \texttt{emu-sv}~\cite{emu-sv}, a statevector emulator simulating the quantum annealing process detailed in Section~\ref{sub:quantum_annealing}. We performed noiseless emulations. The size of the generated subgraphs ranged from 6 to 33 vertices depending on the instance and the hyperparameters.

\begin{figure}[htbp]
\centering
\includegraphics[width=0.45\textwidth]{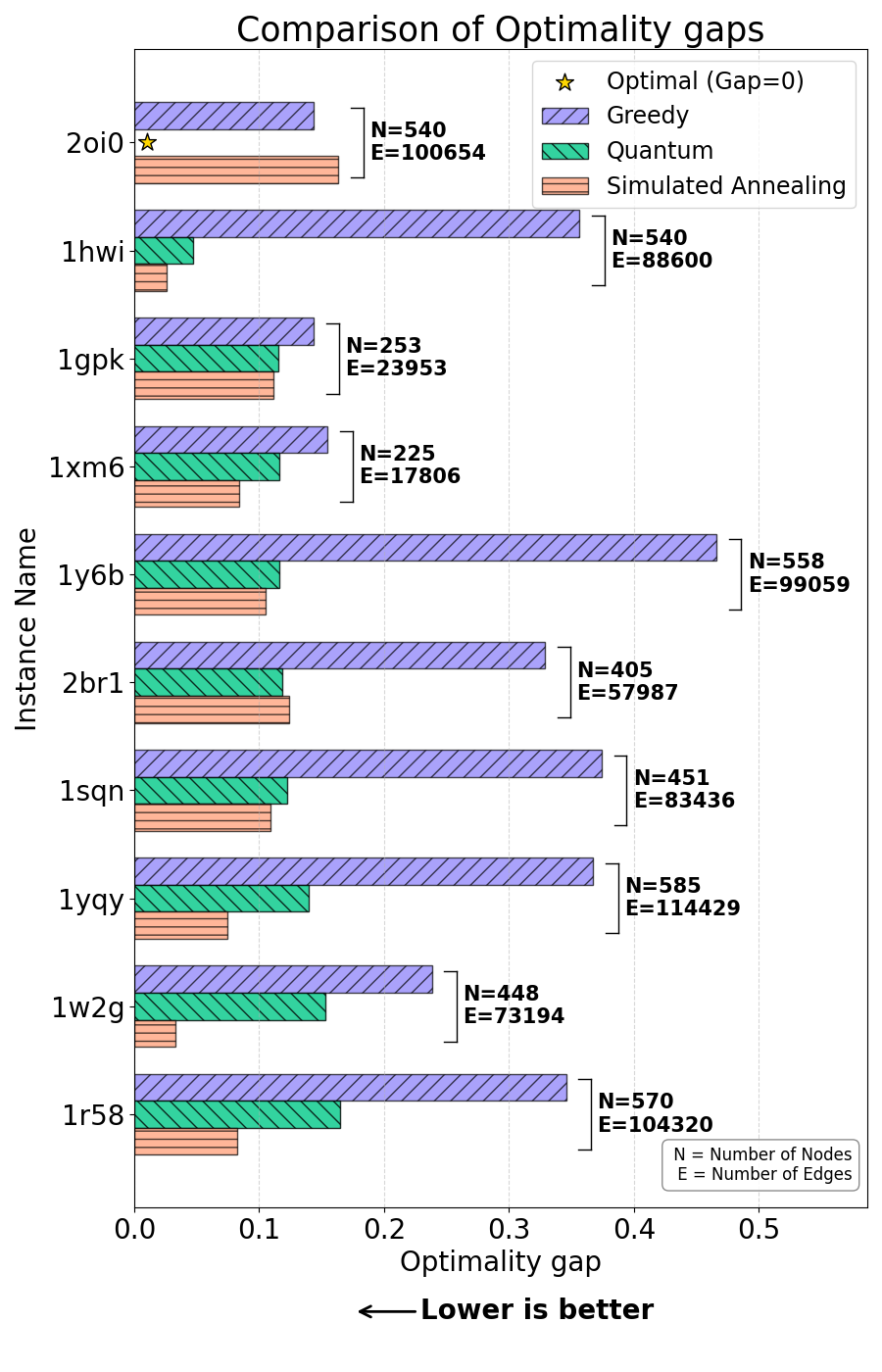}
\caption{Optimality gap for 10 instances of different sizes, TACE-AS (2oi0) being the only instance where the optimal solution was found. Instance name corresponds to its PDB identifier.}
\label{fig:gap}
\end{figure}

The results of our comparative analysis are summarized in Figure~\ref{fig:gap}. Across all 10 instances, the mean optimality gap is $0.109\pm0.049$ for the quantum heuristic, $0.292\pm0.114$ for the greedy baseline, and $0.091 \pm 0.041$ for Simulated Annealing. The quantum heuristic consistently outperforms the greedy method. Simulated Annealing achieves a slightly lower mean gap, which we attribute to the current limitation on subgraph size, discussed further in Section \ref{sec:discussion}.

Notably, the quantum heuristic found the provably optimal solution for the TACE-AS complex (540 nodes, 100,654 edges), matching the exact \texttt{CPLEX} solver. A hyperparameter grid search was conducted on all instances to identify effective configurations. We illustrate this process on the TACE-AS complex, where the optimal solution was reached. The search explored the number of subgraphs $k \in [1,5]$, the number of branches $\ell \in [10,15]$, and the number of solutions retained per step $s \in [1,5]$. The results for this instance are detailed in Table~\ref{tab:grid_search_summary}. The optimal solution was reached with $k=3$, $s=4$, and $\ell=10$.

For practitioners seeking reproducibility without instance-specific tuning, previous work~\cite{cazals2025gadgets} demonstrates that the default configuration $k = 4$, $s = 2$, $\ell = 10$ generally yields competitive results across a variety of graph structures.

\begin{table}[htbp]
\centering
\caption{Performance of the quantum heuristic with varying hyperparameters ($k$, $s$, $\ell$) on the 540-node TACE-AS graph. The optimal MWIS weight is 5.40.}
\label{tab:grid_search_summary}
\begin{tabularx}{\columnwidth}{
>{\centering\arraybackslash\hsize=0.6\hsize}X
>{\centering\arraybackslash\hsize=0.6\hsize}X
>{\centering\arraybackslash\hsize=0.6\hsize}X
>{\centering\arraybackslash\hsize=1.1\hsize}X
>{\centering\arraybackslash\hsize=1.2\hsize}X
}
\toprule
\multicolumn{3}{c}{\textbf{Hyperparameters}} & \multicolumn{2}{c}{\textbf{Performance}} \\
\midrule
$\boldsymbol{k}$ & $\boldsymbol{s}$ & $\boldsymbol{\ell}$ & \textbf{MWIS Weight} & \textbf{Approximation Ratio (\%)} \\
\midrule
1 & 2 & 15 & 4.15 & 76.7\% \\
\addlinespace
2 & 3 & 15 & 4.52 & 83.6\% \\
\addlinespace
\textbf{3} & \textbf{4} & \textbf{10} & \textbf{5.40} & \textbf{100.0\%} \\
\textbf{3} & \textbf{4} & \textbf{15} & \textbf{5.40} & \textbf{100.0\%} \\
\addlinespace
4 & 2 & 15 & 4.94 & 91.4\% \\
4 & 3 & 15 & 5.12 & 94.7\% \\
\addlinespace
5 & 2 & 10 & 5.12 & 94.7\% \\
5 & 3 & 15 & 4.94 & 91.4\% \\
\bottomrule
\end{tabularx}
\end{table}

\subsection{Reconstruction and Biological Relevance}

Once a set of interactions is identified by the solver, it contains a set of receptor points $p_i$ and their corresponding ligand pharmacophore counterparts $l_i$. Since our model incorporates ligand flexibility via a pre-computed conformational ensemble, the reconstruction process must not only find the optimal orientation but also select the single conformer from this ensemble that best satisfies the geometric constraints.

To achieve this, we perform a search over the set of $M$ available conformers. For each conformer, $C_k$, we consider its specific set of pharmacophore coordinates, $\vec{l}_{i,k}$. We then determine the optimal rigid-body transformation (a rotation matrix $R_k$ and a translation vector $\vec{t}_k$) for that specific conformer by minimizing the sum of squared distances to the target receptor points. The objective function for each conformer is therefore:
\begin{equation}
\min_{R_k, \vec{t}_k} \sum_{i=1}^{N} || (R_k\vec{l}_{i,k} + \vec{t}_k) - \vec{p}_i ||^2
\label{eq:kabsch}
\end{equation}
where $N$ is the number of interaction pairs in the clique. This minimization is solved analytically and efficiently for each conformer using the Kabsch algorithm~\cite{Kabsch1976}.

After this procedure is repeated for all $M$ conformers, we identify the best ligand conformer, $C_{best}$, as the one that yielded the overall minimum value for the objective function. The final docked pose is then generated by applying the corresponding best rotation and translation, $(R_{best}, \vec{t}_{best})$, to the entire set of atomic coordinates of this winning conformer.

To assess the biological relevance of the reconstructed poses, we measure the fraction of native contacts ($F_{\text{nat}}$). A native contact is defined as a pair of heavy atoms---one from the receptor and one from the ligand---separated by at most $4.0$~\AA{} in the crystallographic reference structure. $F_{\text{nat}}$ is then the fraction of these reference contacts that are recovered in the predicted pose:
\begin{equation}
F_{\text{nat}} = \frac{|\mathcal{C}_{\text{pred}} \cap \mathcal{C}_{\text{ref}}|}{|\mathcal{C}_{\text{ref}}|}
\label{eq:fnat}
\end{equation}
Following the CAPRI assessment criteria~\cite{janin2003capri}, $F_{\text{nat}} \geq 0.5$ indicates a high-quality prediction, $0.3$--$0.5$ an acceptable one, $0.1$--$0.3$ a low-quality prediction, and values below $0.1$ are considered incorrect.

We perform this reconstruction for all 10 complexes, using both the quantum heuristic and \texttt{CPLEX} solutions. The results are summarized in Table~\ref{tab:fnat_results}. The mean $F_{\text{nat}}$ is $0.125$ for the quantum heuristic and $0.149$ for \texttt{CPLEX}. Most values fall in the low-quality range ($0.1$--$0.3$), with one instance (1gpk) reaching the acceptable threshold for the quantum solver ($F_{\text{nat}} = 0.320$). The slightly higher mean $F_{\text{nat}}$ obtained with the exact solver is consistent with the expectation that a better solution to the MWIS should, on average, yield more physically relevant contacts, which serves as an indicator of coherence for the underlying graph model.

\begin{table}[htbp]
\centering
\caption{Fraction of native contacts ($F_{\text{nat}}$) recovered by the quantum heuristic and the exact CPLEX solver across all 10 benchmark complexes. Boldface indicates the best $F_{\text{nat}}$ per instance.}
\label{tab:fnat_results}
\begin{tabularx}{\columnwidth}{
>{\arraybackslash\hsize=1.6\hsize}X
>{\centering\arraybackslash\hsize=0.4\hsize}X
>{\centering\arraybackslash\hsize=0.2\hsize}X
>{\centering\arraybackslash\hsize=0.8\hsize}X
>{\centering\arraybackslash\hsize=0.8\hsize}X
}
\toprule
\textbf{Complex} & \textbf{PDB} & \textbf{$|V|$} & \textbf{$F_{\text{nat}}$ Quantum} & \textbf{$F_{\text{nat}}$ CPLEX} \\
\midrule
TACE                      & 2oi0 & 540 & \textbf{0.121} & 0.103 \\
Carbonic anhydrase II     & 1hwi & 540 & 0.053 & \textbf{0.171} \\
Dihydrofolate reductase   & 1gpk & 253 & \textbf{0.320} & 0.120 \\
Retinoic acid receptor    & 1xm6 & 225 & 0.038 & \textbf{0.154} \\
Factor VIIa               & 1y6b & 558 & 0.063 & \textbf{0.105} \\
Aldose reductase          & 2br1 & 405 & 0.210 & \textbf{0.242} \\
Neuraminidase             & 1sqn & 451 & \textbf{0.065} & 0.043 \\
Adenosine deaminase       & 1yqy & 585 & 0.125 & \textbf{0.159} \\
GSK-3$\beta$              & 1w2g & 448 & 0.027 & \textbf{0.247} \\
PDE4B                     & 1r58 & 570 & \textbf{0.229} & 0.145 \\
\midrule
\textbf{Mean}             &      &     & 0.125$ \pm0.098$ & \textbf{0.149}$\pm 0.062$ \\
\bottomrule
\end{tabularx}
\end{table}

\section{DISCUSSION AND OUTLOOK}
\label{sec:discussion}
We presented a complete, end-to-end workflow for solving the graph-based formulation of molecular docking on neutral-atom quantum hardware, and validated it on 10 real-world protein-ligand complexes. The divide-and-conquer heuristic~\cite{cazals2025gadgets} decomposes large molecular interaction graphs, previously too extensive for direct QPU embedding, into series of smaller, tractable sub-problems. On the TACE-AS complex (540 nodes), the quantum heuristic matched the exact \texttt{CPLEX} solver, and consistently outperformed the greedy baseline across all instances. Simulated Annealing achieved a slightly lower mean optimality gap ($0.091$ vs.\ $0.109$), which we attribute to the current limitation on subgraph size inherent to the decomposition: the recursive extraction of unit-disk subgraphs constrains each quantum sub-problem to a small number of vertices, limiting per-step solution quality. Improving the subgraph extraction strategy, for instance through more expressive partitioning schemes or lattice geometries, is a key algorithmic challenge for closing this gap.

The biological relevance of the reconstructed poses, assessed via $F_{\text{nat}}$ (Table~\ref{tab:fnat_results}), reveals that most predictions fall in the low-quality range according to CAPRI criteria. This observation holds for all solvers, including the exact one. The slightly higher mean $F_{\text{nat}}$ obtained with \texttt{CPLEX} ($0.149$ vs.\ $0.125$) tends to confirm the coherence of the graph model.  The moderate $F_{\text{nat}}$ values observed across all solvers (including the exact one) indicate that improving the optimization alone will not be sufficient to achieve accurate docking predictions: the graph model itself must also be enriched to capture a more complete picture of molecular interactions. The pharmacophore-based representation treats interactions in a discrete manner and, more critically, rewards favorable contacts without penalizing physically unfavorable configurations such as steric clashes. Introducing negative weights for incompatible interaction pairs would move the model toward a more balanced, free-energy-like representation. Combining our quantum-based global search with a classical refinement step is another promising direction. Additionally, the current workflow relies on the known crystallographic pose of the ligand to define the binding site and select receptor pharmacophores. Removing this assumption, for instance by using blind docking or pocket prediction methods to define the search region, is necessary for the approach to be applicable in a realistic drug discovery setting.

In summary, this work establishes a credible blueprint for applying neutral-atom quantum optimization to large-scale molecular docking, demonstrating clear advances in scalability while delineating concrete axes of improvement, both algorithmic and in the fidelity of the graph model, that will guide future developments.

\bibliographystyle{unsrt}
\bibliography{biblio}
\end{document}